\documentclass[preprint,aps,draft]{revtex4}
\usepackage{dcolumn}
\usepackage{graphics}
\usepackage{bm}
\def\beq{\begin{equation}}
\def\eeq{\end{equation}}
\def\beqar{\begin{eqnarray}}
\def\eeqar{\end{eqnarray}}

\newcommand{\p}{{\partial}}

\newcommand{\gapprox}{\lower.4ex\hbox{$\;\buildrel
>\over{\scriptstyle\sim}\;$}}
\newcommand{\lapprox}{\lower.4ex\hbox{$\;\buildrel
<\over{\scriptstyle\sim}\;$}}
\newcommand{\ov}{{\overline{ v}}}

\begin{document}
\title{Probability distribution function for self-organization of shear flows}

\author{Eun-jin Kim, Han-Li~Liu$^{1}$, and Johan Anderson}
\affiliation{Department of Applied Mathematics, University of Sheffield, Sheffield, S3 7RH,
U.K.\\
$^1$High Altitude Observatory, National Center for Atmospheric Research, Boulder CO 80307, USA}



\pacs{52.35.Ra, 52.25.Fi, 47.27.De, 47.27.eb, 47.35.Bb}
%

\begin{abstract}

The first prediction of the probability distribution function (PDF)
of self-organized shear flows is presented in a nonlinear diffusion
model where shear flows are generated by a stochastic forcing while
diffusing by a nonlinear eddy diffusivity.
A novel non-perturbative method based on a coherent structure
is utilized for the prediction of the strongly intermittent exponential PDF tails
of the gradient of shear flows.
Numerical simulations using Gaussian forcing not only confirm these
predictions, but also reveal the significant contribution from the PDF
tails with a large population of super-critical gradients. 
The validity of the nonlinear diffusion model
is then examined using a threshold model where eddy diffusivity
is given by discontinuous values,  elucidating an important role
of  relative time scales of relaxation and disturbance in the
determination of the PDFs.

\end{abstract}

\pacs{47.65.+a, 52.25.Fi, 52.35.Mw, 52.35.Ra, 52.55.Dy}
\maketitle

\vspace{0.1cm}
\section{Introduction}
Many important phenomena in nature are often far from equilibrium, strongly driven
by instabilities or by external forces.
Examples are diverse, from forest-fires to
interstellar turbulence, which is constantly stirred by super-nova
explosions. Multi-scale interactions are responsible for inevitably
complex dynamics in these non-equilibrium systems,
a proper understanding and description
of which remains as a significant challenge in classical
physics. As a remarkable consequence of multi-scale interactions, a quasi-equilibrium
state can however be maintained by hovering around a {\it marginal} state by a continuous adjustment
of perturbations to establish a new equilibrium \cite{ITOH03}.
While a small perturbation can be transported by
the excitation of waves around a quiet equilibrium state, the
relaxation of a large deviation can involve ballistic, avalanche-like events
of large amplitude on a short dynamical time scale.
This is an essential feature of the so-called self-organization, or
self-organized criticality (SOC) in a more restricted sense \cite{BAK87}.
In particular, it appears to be a powerful paradigm for understanding
complexity in plasmas, with
a growing body of supporting evidence for self organization from
computer simulations, experiments, and observations in laboratory
and astrophysical plasmas \cite{HILDAGO02,SATTIN06,ASTRO,GARBET,XU05,avalanche,LIUETAL02,CHARBONNEAUETAL01}.

The purpose of this paper is to provide a statistical theory of
self organization, which can perhaps be utilized as an exploratory model
in different contexts. As a concrete example, we consider
a forced shear flow, whose gradient grows until it becomes
unstable according to the stability criterion. For instance, in a
strongly stably stratified
medium, fluctuations on small scales (or internal gravity waves)
will sharpen the structure of a
shear flow $u$ \cite{KIMGRAVITY03,KIMGRAVITY07},
acting as a forcing, until its gradient $\partial_x u = u_x$ exceeds the critical value $u_{xc}$,
set by Richardson criterion $R> R_c = (u_{xc}/{\cal N})^2=1/4$. Here ${\cal N}$
is the buoyancy frequency due to the restoring force (buoyancy) in a stably stratified
medium. Once it becomes unstable, the shear flow will relax
its gradient rapidly and generate turbulence (fluctuations) until
it starts building up again at the expense of fluctuations.
In magnetically confined plasmas, poloidal shear flows (zonal flows) and/or parallel
flows can be generated from drift waves while becoming subject to Kelvin-Helmholtz
type instabilities \cite{ROGERS00,HILDAGO02}.
Although precise physical mechanisms for the generation and damping may differ,
the repetition of growth and damping of shear flow is generic, occurring
in many other systems,
playing a crucial role in  momentum transport, mixing, etc.

We model the essential physics involved in the self-organization alluded to above
by the following one dimensional (1D) nonlinear diffusion equation
for $u_x$ \cite{LIU07},
\begin{equation}
\partial_t u_x =  \partial_{xx} [D(u) u_x] + f\,,
\label{eq1}
\end{equation}
where
\begin{equation}
D(u) = \nu + \beta u_x^2\,.
\label{eq2}
\end{equation}
In Eqs.\ (\ref{eq1})-(\ref{eq2}), $f$ is an external forcing;
$D(u)$ represents the effective diffusion coefficient including both the molecular diffusivity
$\nu$ and nonlinear (eddy) diffusivity capturing relaxation process
for unstable shear flow $|u_x| > u_{xc}$.
A similar quadratic eddy diffusivity has
widely been used in modelling chemical mixing and angular momentum transport
(e.g. in stars and the Sun) although the precise value of parameter $\beta$
has been controversial, often adjusted in an attempt to reproduce
observational data \cite{PINS89}.

Since the nonlinear diffusion in Eq. (\ref{eq2}) becomes important
for large gradient $|u_x|> \sqrt{\nu/\beta}$, inhibiting
further increase in the gradient, the critical gradient is roughly
$u_{xc} \sim  \sqrt{\nu/\beta}$.
Due to the relaxation of the gradient above this critical value,
a {\it naive} expectation is that the value of gradient mostly remains subcritical.
This would however be the case only when a relaxation time
is sufficiently short compared with the characteristic time scale of the
perturbation (forcing). In the realistic situation with continuous
perturbation (a stochastic forcing),
there will be a broad distribution of the gradient,
some values of which may exceed far above its critical value.
The key quantity to be determined is thus
{\it the probability distribution function (PDF) of the gradient,
rather than its average value}. In the following, we provide the
first prediction of the PDF tails of the gradient by analytical
and computational studies. Specifically, we show
that PDFs tails for large $|u_x| > u_{xc}$ are
strongly non-Gaussian (intermittent)
with an exponential scaling $\exp{(-c u_x^4)}$
while near the center for small $|u_x|< u_{xc}$, PDFs
are Gaussian $\exp{(-c u_x^2)}$
($c=$ constant).
We then discuss a  threshold model where $D(u)$ is given
by discontinuous values to examine the validity of the nonlinear
diffusion model, elucidating
a crucial role of relative time scales of relaxation and disturbance
in the PDFs.
Note that there has been a growing interest in statistical analysis of SOC
by using various statistical measures, including
PDFs of avalanches
\cite{XU05,avalanche}.

The remainder of the paper is organized as follows. We present
analytical and numerical results of the PDFs of shear 
in a nonlinear diffusion model in \S 2. The predicted power spectrum 
in this model is provided in \S 3. A threshold model is
investigated in \S 4, with numerical results presented. Section 5 contains
Conclusion.

\section{PDFs of shear in a nonlinear diffusion model}
In this section, we provide analytical prediction and numerical
simulation results of the PDFs of the shear $u_x$ in
a nonlinear diffusion model (\ref{eq1})-(\ref{eq2}),
in particular, showing the agreement on strongly intermittent exponential PDF tails
of $u_x$. 

\subsection{Analytic result}
Since for small $|u_x| \ll u_{xc}$, the forcing is balanced by linear diffusion,
naturally leading to the Gaussian distribution of $u_x$,
we focus on the PDF tails for large value of $|u_x|$
where the cubic nonlinearity becomes important.
In order to incorporate this nonlinear interaction non-perturbatively, 
our key idea is to look for a nonlinear structure that is likely to
be naturally sustained in a system.
One candidate for
such a nonlinear structure is an exact nonlinear solution
$u_x\propto x$ to   Eqs. (\ref{eq1})-(\ref{eq2}) in the
absence of the forcing. 
Due to a stochastic forcing, this structure is then likely to form
in a random fashion with the temporal behaviour governed by $Q(t)$
as $u_x \sim i Q(t) x$.
Note that similar coherent structures (ramps) have also been successfully used in
the prediction of the intermittent PDF tails of (positive) velocity gradient
in Burgers turbulence \cite{POLY,GURARIE} that agree with numerical results.
The PDF of $u_x $ then
becomes equivalent to that of $Q(t)$, which satisfies:
\begin{equation}
\partial_t Q = -\beta Q^3 +  g\,,
\label{eq3}
\end{equation}
where $g$ is the time dependent part of the forcing with the spatial
profile $\propto x$.
In the case of a temporally short-correlated forcing
\begin{equation}
\langle g(t_1) g(t_2) \rangle =  \delta (t_1-t_2)  G\,,
\label{eq4}
\end{equation}
the Fokker-Planck equation for the PDF of $Q$ can be derived by using a standard
technique \cite{ZINN}. To this end, we introduce the generating function $Z(\lambda, t)
= e^{-i \lambda Q}$ to obtain
\begin{equation}
\partial_t \langle Z \rangle =
\beta \lambda  \partial_{\lambda \lambda \lambda}\langle Z \rangle
- \lambda^2 G \langle Z \rangle\,,
\label{eq5}
\end{equation}
where $\langle Z \rangle = \int dQ P(Q,t) e^{ - i \lambda Q} =
{\tilde{P}}(\lambda,t)$ is $Z$ averaged over the forcing $g$, which
is equal to Fourier transform of $P(Q, t)$.
The Fourier transform of Eq. (\ref{eq5}) then gives us
the evolution equation for $P$ as:
\begin{equation}
\partial_t P(Q,t) =
\beta \partial_Q [Q^3 P]
+  G \partial_{QQ} P \,.\\
\label{eq6}
\end{equation}
A stationary solution of Eq. (\ref{eq6}) can easily be found to
be $P(Q,t) \sim P_0 \exp{[- \beta Q^4/(4G)]}$, leading to
\begin{equation}
P(u_x;x,t) \sim P_0 \exp{[-\beta u_x^4/4G]} \,.
\label{eq7}
\end{equation}
The PDFs tails in Eq. (\ref{eq7}) are non-Gaussian, intermittent
with the exponential scaling of
$P \sim \exp{(-\beta u_x^4/4G)}$, and
symmetric under the
reflection $x \to -x$ (unlike Burgers turbulence \cite{POLY,GURARIE,BURGERS2} which
is anti-symmetric).
This exponential tail is one manifestation of intermittency caused by
a coherent structure.

To complement the Fokker-Plank approach, it is instructive to
consider an alternative non-perturbative method based on a path
integral formulation \cite{GURARIE,KD02,KA08,AK08}.
A key concept in this method is similar to what was alluded in our
Fokker-Plank approach in that a temporally localized, nontrivial vacuum state
with a coherent structure  -- the so-called instanton which maximizes the path integral
-- causes intermittency, contributing to
the PDF tails. Main steps involved in the  computation of the
PDFs by the instanton method are as follows .
First, we express the PDFs of the
velocity gradient $u_x = v$ to take the value of
$A$ [$P(A)$] in terms of a path integral:
\begin{equation}
P(A) = \int d\lambda e^{i\lambda A - S_\lambda}\,,
\label{eq12}
\end{equation}
where the effective action $S_\lambda$ is given by
\begin{eqnarray}
S_\lambda
& = &-i \int dx dt {\overline{v}} [ \partial_t v -
\partial_{xx} ( \nu + \beta v^2)v]
\nonumber \\
&+& {1\over 2} \int dx dy dt {\overline{v}} (x,t) \kappa(x-y) {\overline{v}}(y,t)
\nonumber \\
&+& i \lambda \int dx dt v(x) \delta(x-x_0) \delta (t)\,.
\label{eq13}
\end{eqnarray}
Here, ${\overline{v}}$ is the conjugate variable to $v=u_x$. By
using the ansatz for temporally localized solutions $v = F(t) \phi$
and ${\overline{v}} =\mu(t) {\overline{\phi}}$
in Eq. (\ref{eq12}) and then by maximizing the effective action
$S_\lambda$  with
respect to $F$ and $\mu$, we obtain the equations for $F$ and $\mu$
(with $\nu=0$) as follows:
\begin{eqnarray}
\partial_t F  - \beta c_2 F^3 = -i c_3 \mu\,, &&
\label{eq15}\\
\partial_t \mu +  3 \beta c_2 F^2 \mu  = -\lambda c_4 
\delta(t) \,, &&
\label{eq16}
\end{eqnarray}
where
\begin{eqnarray}
&&c_1 = \int dx {\overline{\phi}}(x) \phi(x)\,,
\nonumber \\
&&c_1 c_2 = \int dx {\overline{\phi}}(x) \partial_{xx} [\phi(x)^3]\,,
\nonumber \\
&&c_1 c_3 = \int dx dy {\overline{\phi}}(x) \kappa(x-y) {\overline{\phi}}(y)\,,
\nonumber \\
&&c_1 c_4 = {\phi}(x(t=0)) \equiv {\phi(x_0)}\,.
\label{eq17}
\end{eqnarray}
Since instanton $v$ propagates forward in time and its
conjugate variable ${\overline v}$ backward in time while the PDF is computed at $t=0$,
the boundary conditions on $F$ and $\mu$ are: 
\beqar
F(-\infty)= 0\,,
&& \mu(t>0) = 0\,.
\label{eq011}
\eeqar
For $t < 0$, Eqs. (\ref{eq15}) and (\ref{eq16}) give us the equation for $F(t)$ as
\begin{eqnarray}
\p_{tt} F &=& 3 c_2 \beta^2 F^5\,, 
\label{eq17a}
\end{eqnarray}
which can be solved with the boundary conditions 
$F(t=0) = F_0$ and $F(t\to -\infty) =0$ [Eq. (\ref{eq011})];
\begin{eqnarray}
&&F = { F_0 \over (1+2 \beta c_2 F_0^2 t)^{1/2}}\,.\hskip0.5cm
\label{eq19}
\end{eqnarray}
To find the value of $F_0$, we integrate Eq. (\ref{eq16}) for an infinitesimal
time interval $t=[-\epsilon, \epsilon]$ by using Eq. (\ref{eq011}) as: 
\beq
\mu(-\epsilon) = \mu(0) = \lambda c_4\,,
\label{eq000}
\eeq
and substitute Eq. (\ref{eq000}) and $\p_t F = -\beta c_2 F^3$ [from Eq. (\ref{eq19})]
in Eq. (\ref{eq15}) to obtain 
\beq
F_0^{3} =   {i c_3 c_4 \lambda \over 2  \beta c_2} \equiv q \lambda\,,
\label{eq19a}
\eeq
where $q=i  c_3 c_4/2 c_2 \beta$. 

The determination of the PDFs of $P(A)$ now requires a few more 
steps. First, we evaluate $S_\lambda$ in Eq. (\ref{eq12})
by using Eqs. (\ref{eq19}), (\ref{eq000}) and (\ref{eq19a}):
\beq
S_\lambda = Q \lambda^{4/3}\,,
\label{eq191}
\eeq
where $Q = 3i(c_1 c_4) q^{1/3}/4$.
The next step is to find the PDF tails 
by computing the $\lambda$ integral (\ref{eq12}) in the limit of large $\lambda$.
To this end, we substitute Eq. (\ref{eq191}) into Eq. (\ref{eq12}) and 
approximately evaluate $\lambda$ integral as $\int d \lambda   e^{i \lambda A - S_\lambda} \equiv
\int d \lambda e^{-G(\lambda)} \sim e^{-G(\lambda_0)}$,
where $G(\lambda) = -i\lambda A + S_\lambda$ and 
$\lambda_0 = ({3 i  A /4 Q})^3$
is a saddle-point
which minimizes $G(\lambda)$.
Therefore, $P(A)$
in Eq. (\ref{eq12})
becomes 
\begin{eqnarray}
&&P(A) \propto e^{ -\xi \left({A\over \phi(x_0)}\right)^4},\hskip0.3cm
\xi = {\beta  \over 2}
\left|{(c_2 c_1) c_1 \over (c_3 c_1)}\right|\,.
\label{eq21}
\end{eqnarray}
Equation (\ref{eq21}) is the exponential tail of  PDF of $u_x$ to take the value
of $A$, with the same exponential scaling as in 
(\ref{eq7}).
Of importance to notice is that the result (\ref{eq21}) follows from 
the order of the highest cubic
nonlinearity in Eq. (\ref{eq1}), being independent of the precise form of
the spatial structure of $\phi$ and ${\overline{\phi}}$, which has
not been specified yet.
The latter however plays a crucial role in the determination of the overall
amplitude of the PDFs through the coefficient $\xi$ [see Eq. (\ref{eq21})].
Fortunately, the form of $\phi \propto x $ and ${\overline{\phi}} \propto x^3$
(i.e. the exact nonlinear solutions to $v$ and ${\overline{v}}$)
can be inferred from the instanton equations
$\partial_t v - \partial_{xx}(\beta v^3) = -i \int dy \kappa(x-y) \overline{v}(y,t)$ and
$\partial_t \ov + \partial_{xx}(3 \beta v^2 \overline{v}) = -\lambda v(x_0) \delta(t)$,
obtained
by minimizing $S_\lambda$ with respect to $v$ and ${\overline{v}}$,
and by then using $\kappa(x-y) \sim \kappa_0 [1 - (x-y)^2/2 + \cdot\cdot\cdot]
= \kappa_0[ 1 + xy + \cdot\cdot\cdot]$. The use of
$\phi \propto x $ and ${\overline{\phi}} \propto x^3$ in Eqs. (\ref{eq17}) and
(\ref{eq21})
then gives $|c_2 c_1/c_3| \sim 6/\kappa_0$, and thus $\xi \sim  3 (\beta/\kappa_0) $.

To summarize, both Fokker-Planck and instanton methods, 
based on the key
idea that the PDFs tails are caused by a coherent structure,
give us the strongly intermittent PDFs tails of shear gradient $u_x$,
 with the same exponential scalings $\exp{(-c u_x^4)}$ 
($c=$ constant) [see Eqs. (\ref{eq7}) and (\ref{eq21})].
It is interesting to compare these results with the right PDFs of the velocity
gradient in Burgers turbulence, which was predicted to be exponential  
with a different exponent [i.e., $\exp{(-c u_x^3)}$] 
due to ramp-like coherent structures ($u \propto x$) \cite{GURARIE} 
(followed by numerical verification). 
This scaling with the different
exponent basically results from the quadratic highest nonlinear interaction
in Burgers turbulence, different from the cubic highest 
interaction in our model (\ref{eq1})-(\ref{eq2}) (see \cite{KA08} for more
details).
In plasma turbulence, exponential PDFs tails of various fluxes have
been theoretically predicted without numerical confirmation
(e.g. see \cite{KD02,KA08,AK08}).
Nevertheless, it is very interesting that these exponential scalings have often been
observed in the tails of fluxes in laboratory plasmas
(e.g. see Refs. \cite{MRD08,ZWEBEN07}).


\subsection{Numerical Results}

To test our analytical prediction (\ref{eq7}) and (\ref{eq21}), we perform
direct numerical simulations by numerically integrating
Eqs. (\ref{eq1}) and (\ref{eq2}) using method outlined in
\cite{LIU07}. To briefly recap, we use finite difference method to solve (\ref{eq1}). The
spatial discretization is second order accurate and the time integration uses Euler-Maruyama
method. Adaptive time stepping is also used for numerical stability of the diffusion term.
For each step of the simulations, the Gaussian noise is produced using the
Box-Muller method~\cite{PRESS},  which gives 
homogeneous, and temporally short-correlated forcing $f$ in Eq. (\ref{eq1})
with the power spectrum $F(k)$:
\begin{equation}
\langle f(k_1, t_1) f(k_2, t_2) \rangle =  \delta (t_1-t_2)
\delta (k_1 + k_2) F(k) \,.
\label{eq8}
\end{equation}
The results for the PDFs of $u_x$, $P(u_x)$,
for a white-noise $F(k) = k^0$ are shown  by the solid line in Fig.~1
for the values of parameters $\nu = 6 \times 10^{-3}$ and $\beta = 6.25 \times 10^{-3}$.
It can clearly be seen that the
PDF is Gaussian near the center but becomes exponential $\exp{(- c u_x^4)}$
in the tails ($c=$ constant). These exponential tails agree perfectly with our
theoretical prediction (\ref{eq7}). To highlight this,
the dotted and dashed lines in Fig. 1 are fits to a Gaussian and to $\exp{(- c u_x^4)}$,
respectively. The cross-over between these two regimes occurs
approximately at the expected critical gradient of $u_{xc} \simeq
\sqrt{\nu/\beta}= 0.98$. The mean value of $|u_x|$ is found to be
smaller than this, with the value about $0.59$.
However, there is yet a significant probability of 20\% of
super-critical gradient $|u_x|>|u_{xc}|$ from the PDF tails.
The intermittent occurrence of super-critical
gradients can be appreciated from the profile of $u_x$ plotted in Fig.~2,
which exhibits a bursty of large gradients.
Reflectional symmetry of the PDF
is also seen in Figs.~1 and 2.

While mathematically, the PDF tails $\exp{(-c u_x^4)}$
result from the highest cubic nonlinearity in the equation for $u_x$ (\ref{eq1}),
physically, they are due to the feedback of
shear on turbulence when it becomes unstable.
That is, while shear is generated by turbulence [modelled by
the forcing $f$ in Eq. (\ref{eq1})], it feeds back on turbulence,
limiting its own growth, thereby reducing the PDF tails
below the Gaussian prediction (see Fig. 1).
We have confirmed that these exponential PDFs tails are robust features
by using  different power spectra  $F(k) \sim k^{-1}$ and $k^{-2}$.

%
%

\section{Power spectrum in a nonlinear diffusion model}
One of the main interests in the previous studies of self-organization (or SOC)
has been power spectrum.
It is thus interesting to examine what prediction can be made on the power spectra
in our model. To this end, we compute
the PDFs of $u_x(k)$, i.e. $P(k,t)$
by observing that the evolution of $k$ mode
involves the cubic nonlinearity due to nonlinear diffusion
in Eq. (\ref{eq2}) of the form $\int dk_1 dk_2 u_x(k_1) u_x(k_2)
u_x( k-k_1-k_2)$. We approximate the latter as $|u_x(k)|^2 u_x(k)$ by
keeping only the dominant coherent interaction (which can be justified
for a narrow spectrum), and rewrite Eq. (\ref{eq1}) as follows:
\begin{equation}
\partial_t u_x \simeq \beta k^2 u_x^3 -\nu k^2 u_x + f\,.
\label{eq9}
\end{equation}
The Fokker-Planck equation for the $P(u_x;k,t)$ can
be obtained as previously, from which a stationary PDF follows as:
\begin{equation}
P(u_x;k,t)= P_0(k) e^{-k^{2+r} u_x^2 ( k^2 u_x^2/2 + \nu) }\,,
\label{eq10}
\end{equation}
where $P_0(k) = 1/\int du_x P(u_x;t)$ is the normalization constant,
and the power spectrum $F(k) = k^{-r}$ is used.
In the linear case, it is easy to see that the power spectrum
$p(k) = \langle |u_x(k)|^2 \rangle =  \int P(u_x;t) |u_x(k)|^2 \propto k^{-r-2}$.
On the other hand, in a strongly nonlinear case,
we find that
\begin{equation}
p(k) \propto k^{-(2+r/2)}\,.
\label{eq11}
\end{equation}
Remarkably, the prediction (\ref{eq11}) agrees very well with the numerical
results shown in Liu \cite{LIU07}. In particular, in the case of the red noise
with $r=2$, Eq. (\ref{eq11}) predicts $k^{-3}$ spectrum, with
a better agreement with numerical result than the prediction
($k^{-3.5}$) from the renormalization theory!

\section{Threshold model}
Our results highlight the importance of
the statistical description of self-organization.
In particular, the population of super-critical
gradients as well as the form of PDF tails can depend on the relative time
scales between disturbance (i.e. forcing) and relaxation.
To show this,
we consider a threshold model where the nonlinear diffusion $D(u)$ in Eq. (\ref{eq1}) is given
by the two discrete values as
$$
D(u) = \cases { \nu, & for $|u_x|<u_{xc}$; \cr
                V~~ (\gg \nu), & for $|u_x|>u_{xc}$. \cr }
$$
Here $\nu$ is molecular diffusivity while
$V$ represents a large diffusion due to avalanche-like events
which efficiently relax super-critical gradients \cite{LIU07}.
To investigate
the extreme limit where the relaxation rapidly occurs on the shortest time scale,
we numerically solve Eq. (\ref{eq1}) by using this discrete $D(u)$ and
by applying forcing when $|u_x|$ is less than
$u_{xc}$ everywhere in the domain
in order to ensure that
the relaxation occurs much faster than the disturbance.
Note that a similar method was used,
for example, in the SOC solar flare models \cite{LIUETAL02,CHARBONNEAUETAL01}.
Numerical simulation results using $\nu = 6 \times 10^{-3}$,
$V =4.5 \times 10^{-2}$, and $u_{xc}=2$ are plotted in Fig.~3,
which shows that only 0.24\% shear is
super-critical.
This is much less than 20\% found in Fig.~1 in the nonlinear diffusion model,
and is due to rapid relaxation by a large diffusion $V$
for $|u_x|/u_{xc}>1$. The resulting PDFs for this
gradient $|u_x|/u_{xc}>1$ are Gaussian
as can be seen in Fig. 3 since
the diffusion in Eq. (\ref{eq1}) is essentially linear.
In comparison, the Gaussian PDFs near the center
for small $|u_x|/u_{xc}< 0.34$ results from
small fluctuations which satisfy Gaussian statistics.
What is very interesting is that there is a window of
piece-wise exponentials $\exp{(-c u_x^4)}$
between these two Gaussian PDFs for
the gradient $0.34 <|u_x|/u_{xc}<1$,
with a significant population 30\%.
This exponential PDFs are similar to
those found in the nonlinear diffusion case for $|u_x|/u_{xc}>1$
although the exact values of $|u_x|/u_{xc}$ for the exponential PDFs are not identical.
Therefore, these results indicate that the nonlinear diffusion can be a reasonable
approximation for a  certain range of the shear values and
relaxation time scales.

We have also performed the simulation by
applying both the forcing and diffusive relaxation simultaneously
to make disturbance time sufficiently short.
The resulting PDFs are found to be Gaussian since the diffusion
[Eq. (\ref{eq1})] in this case is essentially linear except
at $|u_x|= u_{xc}$.
The super-critical population is also higher (19\%) due to the slower relaxation.

\section{Conclusion}
We have presented a statistical theory of self-organization
by utilizing a simplified nonlinear diffusion model for a shear flow and
a widely invoked quadratic eddy diffusivity \cite{LIU07,PINS89}.
Both Fokker-Planck and instanton methods predict the PDF tails
of the exponential form $\exp{(-c u_x^4)}$, with a strong intermittency.
Our numerical simulation using Gaussian
forcing with three different power spectra not only confirm these
predictions, but also reveal the significant contribution from the PDF
tails with a large population of super-critical gradients, which
could play a crucial role.
These results
highlight the importance of the statistical description
of gradients in self-organization, rather than its average value as has conventionally been done.
The validity of the nonlinear diffusion model
was then examined using a threshold model, elucidating an important role
of  relative time scales of relaxation and disturbance,
calling for a care in actual modelling of a particular system.

Our results can have significant implications for
the dynamics and the role of shear flows (e.g. zonal flows) in laboratory, astrophysical and
geophysical plasmas,  which is vital not only
in momentum transport, but also in transporting chemical species and
controling mixing of other quantities (e.g. air pollution, weather control)
\cite{KIMRECENT1,KIMRECENT2}.
Our theory can also provide a useful guide in understanding self-organization in
other disciplines, such as population in environmental dynamics and biology,
forest-fire, and reaction and diffusion in chemistry.
Future work will include specific applications to those systems,
the extension of our model to incorporate the finite correlation time
of the forcing and a non-diffusive flux \cite{HAHM95}, and
the investigation of the joint PDFs of fluctuations and mean gradients
in a consistent way.
Note that an initial attempt to the prediction of the joint PDFs has been
made in the ion temperature gradient turbulence (for magnetically confined
plasmas)
by neglecting the feedback of shear flows on fluctuations \cite{AK08,AK09}. \\

This research was supported by the EPSRC grant EP/D064317/1
and RAS Travel Grant.
The National Center for Atmospheric Research is sponsored by the NSF.

\vfill\eject
\noindent
{\centerline{Figure Captions}}

\bigskip
\noindent
Fig. 1 The solid line is the PDFs from the numerical simulation of a nonlinear diffusion model (\ref{eq2}) for a white noise. The dotted and
  dashed lines are the fits to Gaussian and $\exp{(-c u_x^4)}$ ($c
  =$const).

\bigskip
\noindent
Fig2. The profile of $u_x$ corresponding to Fig.~1.

\bigskip
\noindent
Fig. 3 Solid line is the PDF from a threshold model. Dotted
and dash-dotted-dotted-dotted lines are Gaussian fits;
dashed and dashed-dotted lines are fits to $\exp(-cu_x^4)$
($c=$const).

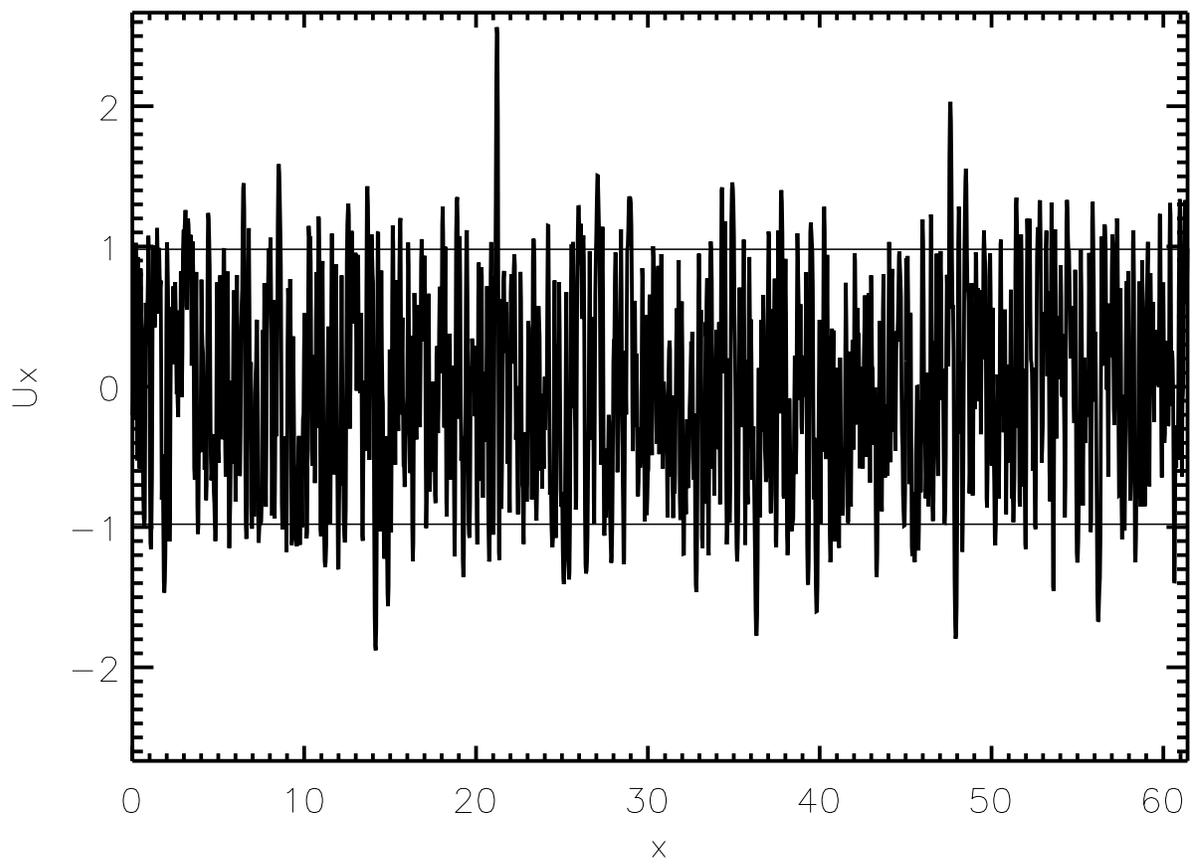
\end{document}